\documentclass[reprint,superscriptaddress,amsmath,amssymb,aps,pra,twocolumn,pra]{revtex4-2}

\usepackage{color, amssymb, graphicx, amsfonts,epstopdf,float}
\usepackage{multirow,amssymb,amsbsy,amsmath,epstopdf}
\usepackage{hyperref}
\usepackage{verbatim}
\usepackage{bm,braket,comment}
\usepackage{bbold}
\usepackage{soul}
\usepackage{subfigure}

\begin{document}

\title{Heralded entangled state generation enhanced by photon addition and subtraction}

\author{Yun-Long Cao}
\affiliation{Laboratory of Quantum Information, University of Science and Technology of China, Hefei 230026, China}
\affiliation{Anhui Province Key Laboratory of Quantum Network, University of Science and Technology of China, Hefei 230026, China}
\affiliation{CAS Center for Excellence in Quantum Information and Quantum Physics, University of Science and Technology of China, Hefei 230026, China}
\affiliation{Hefei National Laboratory, University of Science and Technology of China, Hefei 230088, China}
\author{Xiao-Ye Xu}
\email{xuxiaoye@ustc.edu.cn}
\author{Chuan-Feng Li}
\email{cfli@ustc.edu.cn}
\author{Guang-Can Guo}
\affiliation{Laboratory of Quantum Information, University of Science and Technology of China, Hefei 230026, China}
\affiliation{Anhui Province Key Laboratory of Quantum Network, University of Science and Technology of China, Hefei 230026, China}
\affiliation{CAS Center for Excellence in Quantum Information and Quantum Physics, University of Science and Technology of China, Hefei 230026, China}
\affiliation{Hefei National Laboratory, University of Science and Technology of China, Hefei 230088, China}

\date{\today}

\begin{abstract}
We propose a heralded entanglement generation scheme based on Gaussian sources enhanced by photon addition and subtraction operations. By combining single-mode squeezing, linear interferometers, and conditional photon-number measurements on ancillary modes, our model can probabilistically generate dual-rail encoded Bell, GHZ, and W states. We systematically optimize the squeezing parameters and interferometer settings to maximize both the heralding success probability and the fidelity with the target states. Our results show that photon addition and subtraction significantly enhance the non-classicality of the output states and improve generation performance, while maintaining computational efficiency comparable to single-photon source models. We further analyze the robustness of the scheme under parameter perturbations and find that its performance remains stable under realistic experimental imperfections. This work provides a versatile and experimentally feasible framework for scalable heralded entanglement generation using Gaussian resources with non-Gaussian operations.
\end{abstract}

\maketitle

\section{Introduction}
Quantum entanglement, a hallmark distinction between quantum and classical systems, is widely recognized as a foundational resource for quantum information technologies\,\cite{RevModPhys.81.865,Nielsen_Chuang_2010,RevModPhys.86.419} and has been extensively exploited for tasks such as quantum key distribution, quantum teleportation, one-way quantum computing, and quantum
metrology \,\cite{PhysRevLett.67.661,PhysRevLett.70.1895,PhysRevLett.86.5188,doi:10.1126/science.1104149}. 
The rapid advancement of quantum applications and the ongoing pursuit of quantum supremacy are driving the demand for large-scale quantum resources, thereby motivating the development of entangled states in both high-dimensional and many-body regimes\,\cite{Erhard2020,PhysRevX.8.021012,RevModPhys.84.777}.
As a result, the efficient generation of such states has emerged as a pivotal challenge in advancing quantum technologies.

In photonic quantum information, Gaussian states or Gaussian operations are of particular interest\,\cite{RevModPhys.84.621,doi:10.1142/S1230161214400010}. They are relatively easy to prepare and highly stable for scaling up the experimental system. 
However, Gaussian states and operations are fundamentally limited, as they cannot directly produce certain non-classical states or maximally entangled states that are useful for showing quantum advantage\,\cite{PhysRevLett.109.230503,PhysRevA.97.062337,PhysRevA.82.052341}. 
To overcome this limit, the non-Gaussians operation such as conditional measurements, photon addition and subtraction are introduced to enable the probabilistic generation via post-selection and heralding\,\cite{PhysRevA.100.052301,PhysRevA.59.1658,doi:10.1126/science.1122858,PhysRevA.80.053822,PhysRevA.86.012328}. 
This method inevitably suffers from rapidly decreasing success probabilities and increasing system complexity when the scale of entanglement expands. Optimizing both the success rate and the quality of entanglement is therefore of crucial importance for realizing quantum applications on optical platforms.

Recent studies have explored several approaches to probabilistic entanglement generation. Some rely on single-photon sources based on quantum dots or on heralded parametric down-conversion processes\,\cite{PhysRevResearch.3.043031,PhysRevA.102.012604,PhysRevLett.132.130603,PhysRevLett.132.130604,PhysRevLett.125.110506}. 
Using Gaussian states alone to probabilistically prepare photonic entangled states\,\cite{doi:10.1126/science.aay4354,PhysRevA.100.052301} have been explored, while subsequent work incorporates additional single photons to improve the efficiency of state preparation\,\cite{PhysRevA.109.023717}. 
Using non-Fock bases to encode entangled states like cat-state basis\,\cite{Sychev2017,PhysRevA.99.053816} or using heralded states generated from Gaussian sources\,\cite{PhysRevApplied.17.034071} as inputs are also feasible directions for efficient generation of entanglement. 
These approaches can be efficiently integrated with emerging experimental platforms such as large-scale photonic integrated circuits\,\cite{Arrazola2021,Shekhar2024} and Gaussian boson sampling\,\cite{PhysRevLett.119.170501}, thereby offering new models and promising directions for future quantum technologies.

In this work, we employ Gaussian states and photon-number detection as resources to extend heralded state generation toward entangled states, including Bell states, GHZ states, and W states, encoded in Fock basis. This work integrates the approaches of continuous-variable and discrete-variable schemes in the field of heralded entanglement generation, thereby extending the applicability and potential applications of this method across different physical platforms. By introducing heralded photon addition and subtraction, we further enhance the non-Gaussianity of the system, aiming to achieve superior performance in entanglement generation.

\section{Background}

A composite system defined on the Hilbert space $H$ is a tensor product of the corresponding subsystem spaces $H=\otimes_{l=1}^n H_l$, the total state of the system 
\begin{equation}
	\ket{\psi}=\sum_{i_1,\ldots,i_n} c_{i_1,\ldots,i_n} \ket{i_1}\otimes\ket{i_2}\otimes\cdots\otimes\ket{i_n}
\end{equation}
is separable when it can be described as a product of states in the subsystem spaces \(\ket{\psi}=\ket{\psi_1}\otimes\ket{\psi_2}\otimes\cdots\otimes\ket{\psi_n}\), and the states that are not separable are called entangled. In the bipartite system  with  two-dimensional local subspaces there are the maximally entangled states called Bell states, 
\begin{eqnarray}
	&&\ket{\psi^\pm}=\frac{1}{\sqrt{2}}(\ket{0}\ket{1}\pm\ket{1}\ket{0}),\notag\\
    &&\ket{\phi^\pm}=\frac{1}{\sqrt{2}}(\ket{0}\ket{0}\pm\ket{1}\ket{1}),
\end{eqnarray}
which are well known for the nonlocal inequality test\,\cite{PhysRevLett.23.880}. In multipartite systems, the structure of entanglement becomes significantly more intricate, and there may exist inequivalent forms. For three qubits, there  are two inequal maximal entangled states, Greenberger–Horne–Zeilinger (GHZ) \(\ket{\mathrm{GHZ}}=\frac{1}{\sqrt{2}}(\ket{000}+\ket{111})\) and W \(\ket{\mathrm{W}}=\frac{1}{\sqrt{3}}(\ket{001}+\ket{010}+\ket{100})\) states.
These states cannot be generated from the separable states by local operations and classical communication (LOCC), making their efficient generation a challenge. 

The problems involving multi-mode Gaussian states are typically formulated within continuous-variable (CV)  quantum systems, where information is encoded in the quadratures of bosonic modes rather than in discrete two-level systems\,\cite{RevModPhys.84.621}. An N-mode bosonic system can be described by the quadrature operators of each mode defined in terms of creation and annihilation operators as 
\begin{equation}
	\hat{q}_i = \frac{\hat{a}_i+\hat{a}^\dagger_i}{\sqrt{2\hbar}},\quad
	\hat{q}_i = \frac{-i(\hat{a}_i-\hat{a}^\dagger_i)}{\sqrt{2\hbar}} .
\end{equation}
We define an operator vector \(\bm{x}=(q_1,p_1,...,q_N,p_N)\). The Gaussian states can be completely characterized by the first and second moments of the operator vector. The first moment is called the
displacement vector \(\bar{\bm{x}}=\langle \hat{\bm{x}} \rangle\) and the second moment is called the covariance matrix \(\bm{V}\) with elements \(V_{ij}=\tfrac{1}{2}\langle \hat{x}_i\hat{x}_j+\hat{x}_j\hat{x}_i \rangle-\langle\hat{x}_i\rangle \langle\hat{x}_j\rangle \).  The Wigner function provides a quasi-probability representation of quantum states in phase space and completely characterizes the CV quantum states. Gaussian states are named because their Wigner functions take the form: 
\begin{equation}
	W_G(\bm{x}) =
	\frac{\exp\!\left[-\tfrac{1}{2}(\bm{x}-\bm{\bar{x}})^T \bm{V}^{-1}(\bm{x}-\bm{\bar{x}})\right]}
	{(2\pi)^N \sqrt{\det\bm{V}}} .
\end{equation}

Gaussian operations transform Gaussian states into Gaussian states. The unitary Gaussian operation generated from the Hamiltonian that is second-order polynomial of the mode operators can be described by a symplectic transformation $S$ and a displacement $d$ in terms of the quadrature operators. The displacement vector and the covariance matrix transform as
\begin{equation}
	\bm{\bar{x}}\rightarrow \bm{S}\bm{\bar{x}}+\bm{d}, \quad
	V\rightarrow \bm{S}\bm{V}\bm{S}^T.
\end{equation}
Gaussian operations, including displacements, squeezing and linear mode transformation can be implemented using optical devices, which offers a practical advantage for experimental realization. The partial trace is a nonunitary Gaussian operation that removes degrees of freedom from the phase space description. The reduced state remains Gaussian and is characterized by the displacement vector and the covariance matrix obtained after removing the excluded modes. This property ensures that the description remains within the Gaussian formalism even for conditional states of subsystems\,\cite{PRXQuantum.2.030204}. 

Performing local photon-number measurements from a multi-mode Gaussian state yields the probability distribution on the Fock basis\,\cite{PhysRevLett.119.170501,PhysRevA.100.032326}. For a state characterized by the covariance matrix \(\bm{V}\) and a zero displacement vector, the probability of a specific measurement outcome \(\hat{\bm{n}}=\otimes_j^M\ket{n_j}\bra{n_j}\) is
\begin{equation}
	P(n) = \frac{\mathrm{Haf}(A_S)}{\bm{n}!\sqrt{|V_Q|}},
\end{equation}
where 
\begin{equation}
	V_Q = V+I_{2M}/2,
\end{equation}
\begin{equation}
	A = \begin{bmatrix}
		0 & I_{2M} \\
		I_{2M} & 0
	\end{bmatrix}(I_{2M}-V_Q^{-1}),
\end{equation}
and \(A_S\) is a submatrix from A that the rows and columns \(S\) remain depending on the measurement outcome. The Hafnian is a matrix function summing over all perfect matching permutations. Coherent contributions are present in the distribution when the displacement vector is non-zero. The probability is
\begin{equation}
\begin{split}
	P(n)= &\frac{\mathrm{exp}(-\tfrac{1}{2}\bm{x}^{\dagger}V_Q^{-1}\bm{x})}{\bm{n}!\sqrt{V_Q}}(\mathrm{Haf}(A_S)+ \\
	&\sum_{j_1,j_2,j_1\neq j_2}F_{j_1}F_{j_2}\mathrm{Haf}(A_{S-\{j_1,j_2\}})+ \\
	&...+\prod_j^{2M} F_j),
\end{split}
\end{equation}
where \(\bm{F}=\bm{x}^{\dagger}V_Q^{-1}\). Because of the computational complexity of Hafnian in the \#P class, Gaussian boson sampling has emerged as a leading platform for demonstrating quantum computational advantage.

Post-selection is a probabilistic strategy that obtains the desired outcome by conditioning on specific measurement outcomes while discarding others. For instance, photon pair generated in spontaneous parametric down-conversion (SPDC) sources are probabilistic. Only when coincidence counts are registered at the detectors can the desired entangled states or single photons be extracted. Heralding is a conditional method based on the measurements performed on ancillary modes. The detection of a specific outcome signals the presence of the desired states in the remaining modes, enabling the heralded states to serve as deterministic resources in next steps, such as heralded single photons from SPDC and single-mode nonlinear operations. For a pure state, when a measurement outcome \(\bm{h}\) is obtained on the ancillary modes with probability \(p\), the total state can be decomposed into the following form,
\begin{equation}
	\ket{\psi} = \sqrt{p}\ket{\phi}\ket{\bm{h}}+\sqrt{1-p}\ket{R} 
\end{equation} 
while \(\ket{\phi}\) is the output state and \(\ket{R}\) is the rest state discarded by heralding\,\cite{PhysRevResearch.3.043031}. Both the success probability and the quality of the target state are required for the entanglement generation. 
\section{Method}

A heralded Gaussian model can be divided into two essential steps: the preparation of Gaussian states and the measurement on ancillary modes. Experimentally, multimode Gaussian states are typically generated from multiple single-mode squeezed vacuum states combined through linear mode transformations, which can be decomposed into a finite network of two-mode transformations. The heralding state is chosen from the Fock bases of the ancillary modes that are spatially separated from the output modes. When a specific photon-number pattern is detected on the ancillary modes, the output modes are projected onto a conditional state that closely matches the target entangled state, thereby realizing successful heralded generation. 

\begin{figure}[h]
	\centering
	\includegraphics[width=0.23\textwidth]{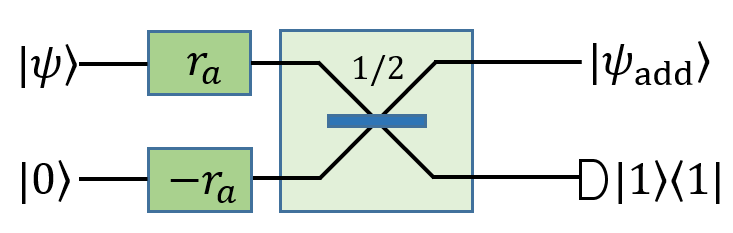}
	\includegraphics[width=0.23\textwidth]{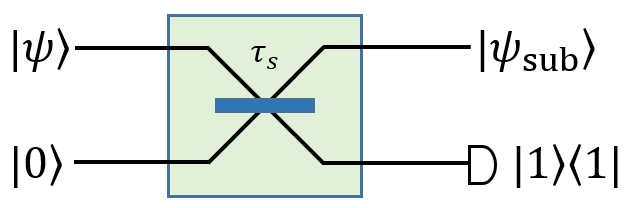}
	\caption{Schematic of photon subtraction using a beam splitter and photon addition using  single-mode squeezing. The squeezing parameter \(r_a\) for photon addition and the beam splitter transmissivity \(\tau_s\) for photon subtraction are set to 0.8 and 0.1 in our model.}
	\label{fig:PA_PS}
\end{figure}
\begin{figure}[h]
	\centering
	\includegraphics[width=0.35\textwidth]{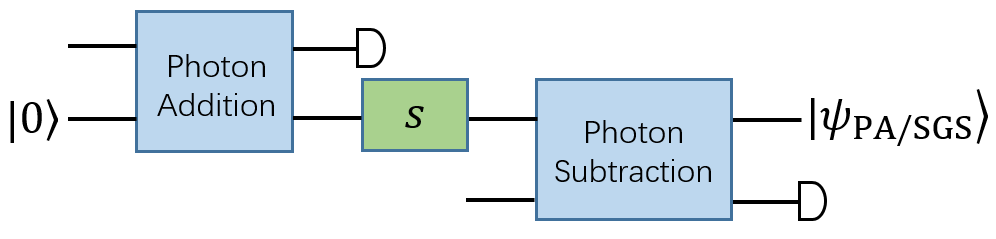}
	\\[4pt]
	\includegraphics[width=0.35\textwidth]{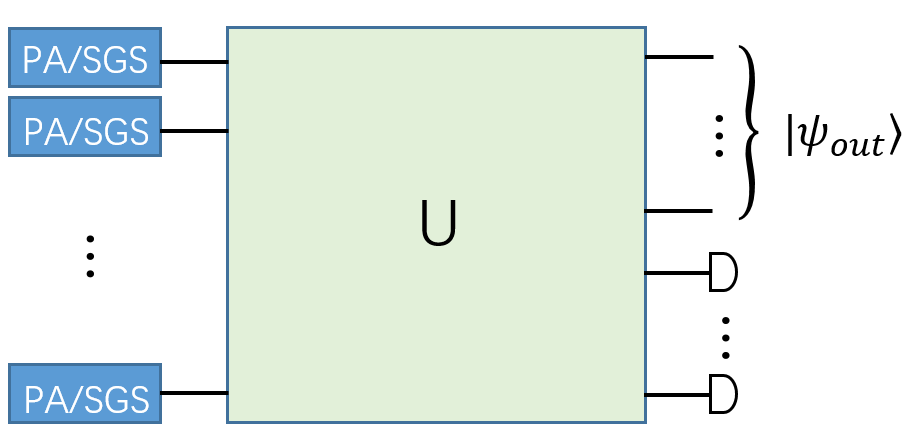}
	\caption{Schematic of heralded entanglement generation model with photon addition/subtraction Gaussian source. The single-mode squeezers S and the linear interferometer U are serving as optimization parameters.}
	\label{fig:Model}
\end{figure}

Moreover, additional modes and Gaussian operations can be introduced to implement photon addition and subtraction, thereby enhancing the non-Gaussianity and non-linearity of the system. A parametric amplifier or a beam splitter combined with a photon-number detector can probabilistically realize photon addition or photon subtraction operation in the optical setup, and these operations can be incorporated into the heralded Gaussian model by enlarging the number of modes. Owing to the similar action of both single photons and Gaussian states with linear mode transformations, our model places photon-addition and subtraction operations adjacent to squeezing operations for obvious impact. 

The target states are encoded in a dual-rail encoding scheme, in which each qubit is represented by occupation in two distinct local modes. For the Bell, GHZ and W state, the corresponding physical states are written as
\begin{align}
	&\ket{\mathrm{Bell}}=\frac{1}{\sqrt{2}}(\ket{1010}+\ket{0101}), \\
	&\ket{\mathrm{GHZ}}=\frac{1}{\sqrt{2}}(\ket{101010}+\ket{010101}), \\
	&\ket{\mathrm{W}}=\frac{1}{\sqrt{3}}(\ket{101001}+\ket{100110}+\ket{011010}),
\end{align}
expressed in the multimode Fock basis. Each basis component of the superposition state is encoded in at least one independent mode, so that the relative phases of the superposition can be directly controlled through single-mode phase rotations, overcoming the limitation that local photon-number measurements cannot reveal phase information.
The quality of the heralded entangled state generation is assessed by both the heralded success probability and the fidelity with the target state. To achieve optimal performance in experiments, we optimize the parameters of single-mode squeezing and the linear interferometer under fixed conditions of photon addition/subtraction and heralding states. 

The cost function is defined as
\begin{equation}
	f(\bm{\xi}) = -w_1\mathrm{ln}(p)-w_2\mathrm{ln}(F)+\epsilon\sum ||\xi_i||^2,
\end{equation}
where \(p\) is the heralding success probability and \(F=|\braket{\psi_{output}|\psi_{target}}|^2\) is the fidelity with the target state, both defined from Fock basis measurement probabilities. The numerical computations in CV quantum system are performed using The Walrus package\,\cite{Gupt2019}. 

Compared with single-photon-source models, in which evaluating conditional probabilities requires summing over all measurement outcomes corresponding to successful heralding, the closure of Gaussian states under partial trace reduces the computational task to evaluating the Hafnian of the reduced state.

\begin{figure}[h]
	\centering
	\includegraphics[width=80mm]{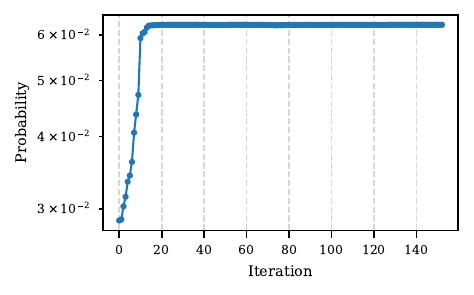}
    \\[4pt]
	\includegraphics[width=80mm]{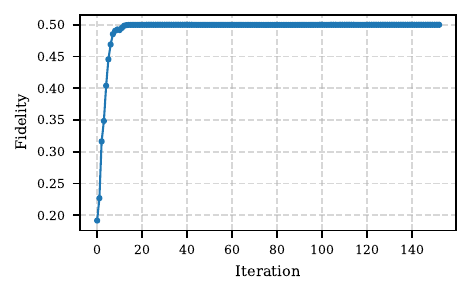}
	\caption{Probability and fidelity of the Bell state generated as a function of optimal steps. The weights \(w_1\), \(w_2\) and \(\epsilon\) are set to \(10\), \(1\) and \(10^{-4}\).}
	\label{fig:trace_1}
\end{figure}

The schematic of the proposed heralded entanglement generation model is illustrated in Fig.~\ref{fig:PA_PS} and \ref{fig:Model}. The Gaussian source is composed of a basic single-mode squeezed state combined with heralded photon addition and subtraction. Photon addition is realized by preparing a two-mode squeezing state, which can equivalently be implemented using two single-mode squeezers followed by a balanced beam splitter. When a single photon is detected on the ancillary mode, the other mode is conditionally projected onto a state with one additional photon. Similarly, photon subtraction is realized by interfering the modes with vacuum on a beam splitter and detecting a single photon on the ancillary mode, thereby heralding photon subtraction on the other mode. Photon addition is applied before the single-mode squeezing to exploit possible stimulated effects, while photon subtraction is performed after the squeezing to mitigate vacuum-induced errors.

\begin{figure}[h]
	\centering
	\includegraphics[width=80mm]{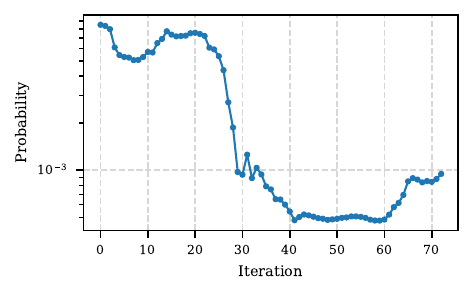}
	\\[4pt]
	\includegraphics[width=80mm]{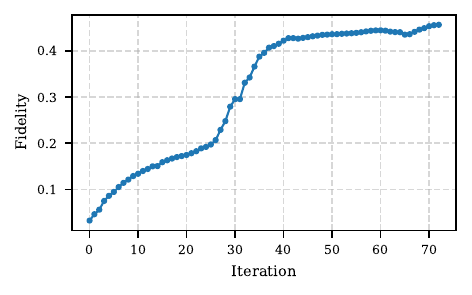}
	\caption{Probability and fidelity of the GHZ state generated as a function of optimal steps. The weights \(w_1\), \(w_2\) and \(\epsilon\) are set to \(10\), \(1\) and \(10^{-4}\).}
	\label{fig:trace_2}
\end{figure}

The linear optical interferometer is constructed by cascading multiple beam splitters and phase shifters, which allows any multimode interferometer SU(n) to be decomposed into a sequence of SU(2) transformations\,\cite{Clements:16}. Heralding measurements are implemented via single-mode photon-number-resolving detections, which project the Gaussian state onto the multimode Fock basis. In this framework, the heralding modes conditionally project the output modes onto entangled states, such as Bell or GHZ states—depending on the parameters and measurement outcomes. Since the vacuum component inherent in Gaussian states limits the nonclassicality of the heralded states, post-selection on the output modes can be employed to exclude vacuum events. 
\section{Results}

\subsection{Optimization Procedure}
Rapid and stable convergence of the success probability and fidelity during the optimization process is essential for the practical applicability and scalability of the model. To assess the computational feasibility of our scheme, we plot the probability and fidelity of the generated entangled states as functions of the optimization steps for different target states and heralding measurements, and compare them with those of single-photon-source models of similar scale. 

\begin{figure*}[t]
	\centering
	\includegraphics[width=55mm]{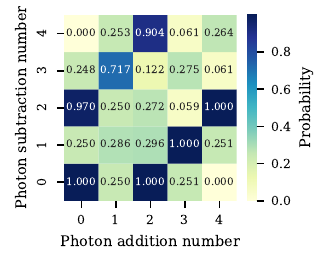}
    \includegraphics[width=55mm]{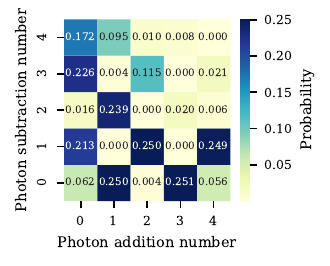}
	\includegraphics[width=55mm]{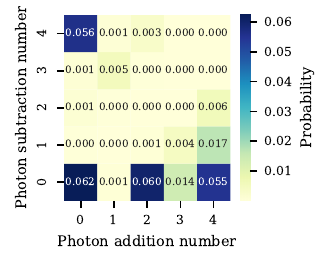}	
	\\[4pt]
	\includegraphics[width=55mm]{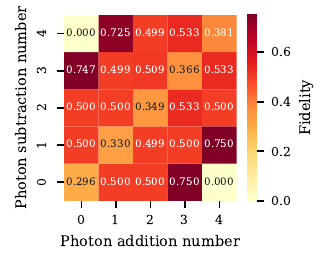}
    \includegraphics[width=55mm]{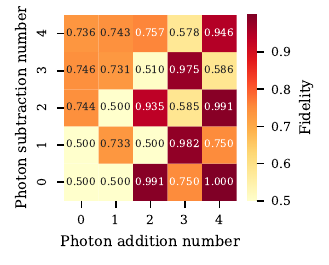}
	\includegraphics[width=55mm]{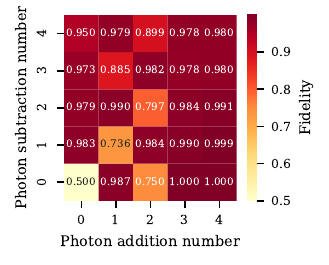}
	\caption{Probabilities (top row) and fidelities (bottom row) of the Bell state generated as a function of photon addition and subtraction for different numbers of heralding modes as 1, 2 and 3 in each column.}
	\label{fig:heatmap_Bell}
\end{figure*}

For Bell state (Fig.~\ref{fig:trace_1}), we encode in two dual-rail qubits across four physical modes. The heralding modes are chosen to all-click events or events in which only one mode is in vacuum, depending on whether the photon number generated by squeezing is even or not.
For comparison with a single-photon-input model of the same scale, we use two heralding modes, and both models converge within \(10^2\). When the model is extended to the generation of GHZ state (Fig.~\ref{fig:trace_2}), encoded in three dual-rail qubits across six physical modes, we set the number of heralding modes to four to match the single-photon input model of comparable scale, which also converges within \(10^2\) iterations. 
These results indicate that our Gaussian-source model exhibits comparable optimization complexity to the single-photon-source model in various heralded entanglement generation tasks, while achieving the stable optimal probability and fidelity. Compared with the Bell and GHZ states, the W state is more difficult to optimize because some of its superposition terms share identical photon occupations in certain physical modes. This partial mode overlap makes the optimization of both probability and fidelity more prone to local optima.

\subsection{Photon Addition and Subtraction}
In the single-photon-source model, the number of input photons and heralding modes constrains the probability and fidelity of heralded entanglement generation\,\cite{PhysRevA.96.043861}, which remains equally important in our model. Since Gaussian states do not conserve photon number, single-photon conditional measurements is unable to completely eliminate the limitations imposed by vacuum and multi-photon components on the output state fidelity. Introducing additional non-Gaussian operations, such as photon addition and subtraction, can improve the performance of entanglement generation. Therefore, selecting the optimal heralding modes and photon addition/subtraction configurations is a key research direction for achieving efficient and high-fidelity entangled state generation. 

Fig.~\ref{fig:heatmap_Bell} shows heat maps of the heralded generation probability and fidelity for Bell states under different configurations. In contrast to their consistently beneficial role in enhancing the nonclassicality of Gaussian states and in the heralded generation of cat states and Gottesman–Kitaev–Preskill (GKP) states\,\cite{PhysRevA.109.023717}, photon addition and subtraction have a more configuration-dependent effect on the heralded generation of dual-rail-encoded entangled states. In some cases, such as addition two photons and no subtraction with two heralding modes, the Gaussian-state parameters can be optimized to achieve a reasonable probability of generating entangled states with high fidelity. In the best configuration, a Bell state with a fidelity of 0.991 can be generated with a probability of 0.004. Given the probabilistic nature of heralded single-photon sources, this efficiency is comparable to that of the single-photon source model. In other configurations, however, the optimization tends to become trapped in local optima or fails to improve probability and fidelity simultaneously. 

The results show that increasing the number of heralding modes generally decreases the generation probability while improving the fidelity, whereas increasing the number of photon additions and subtractions can enhance both the probability and fidelity. Moreover, photon addition proves to be more effective than photon subtraction, as it is more effective at suppressing the contributions of the vacuum and multi-photon components in the output state. When the number of photon addition reaches four and the number of heralding modes exceeds two, the fidelity approaches unity, at which point the Gaussian-source model becomes essentially equivalent to the single-photon-source model affected by Gaussian noise. 

\begin{figure}[h]
	\centering
	\includegraphics[width=60mm]{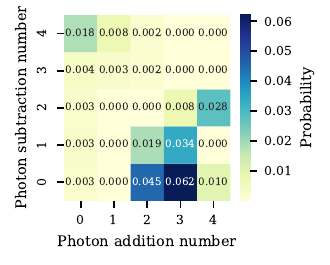}
	\\[4pt]
	\includegraphics[width=60mm]{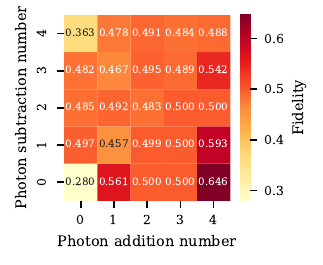}
	\caption{Probability and fidelity of the GHZ state generated as a function of photon addition and subtraction for the numbers of heralding modes as 4.}
	\label{fig:heatmap_GHZ}
\end{figure}

For GHZ states generation, the additional heralding modes with photon additions or subtractions increase the number of modes used. This leads to an exponential growth in the computational complexity for simulation, eventually resulting in the calculation of the Hafnian. For this reason, we present the optimization results only for 4 heralding modes in Fig.~\ref{fig:heatmap_GHZ}. 

The results show that increasing the number of photon additions and subtractions can enhance the probability and fidelity in certain configurations. in configurations such as one added photon and one subtracted photon, the optimization tends toward lower probability and fidelity. Since the computational cost of each optimization step grows exponentially with the number of modes, and the size of the parameter space increases accordingly, directly obtaining GHZ states with both high probability and high fidelity remains challenging through optimization alone.

\subsection{Post-selecting Non-vacuum States}
In the heralded state-generation model considered here, a residual vacuum component in the output modes is unavoidable. This vacuum contribution originates from the intrinsic vacuum dominance of low-gain Gaussian sources and cannot be completely eliminated by a finite number of heralding measurements or photon addition/subtraction operations. As a result, even when the heralding conditions are satisfied, the vacuum component inevitably degrades the fidelity of the generated target states.

\begin{figure}[h]
	\centering
	\includegraphics[width=80mm]{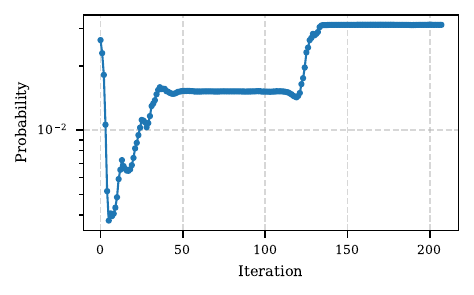}
    \\[4pt]
	\includegraphics[width=80mm]{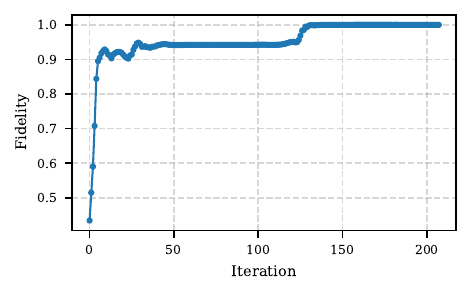}
	\caption{Probability and fidelity of the Bell state generated with post-selecting non-vacuum states as a function of optimal steps. The weights \(w_1\), \(w_2\) and \(\epsilon\) are set to \(10\), \(1\) and \(10^{-4}\).}
	\label{fig:trace_nv_1}
\end{figure}

To mitigate this effect, we adopt a post-selection strategy that removes vacuum contributions at the level of the encoded output state. Following Ref.\,\cite{PhysRevApplied.17.034071}, we assume an ideal on-off quantum nondemolition (QND) measurement on the output modes, which distinguishes vacuum from non-vacuum events without disturbing the encoded dual-rail subspace. If the purpose is solely to verify the generated state, this QND measurement may be replaced in practice by standard threshold detectors, which are readily available experimentally. We post-select only those events in which each pair of physical modes encoding a dual-rail qubit contains at least one photon. Importantly, this procedure does not alter the state-preparation process itself, nor does it introduce additional heralding modes or photon addition/subtraction operations; instead, it restricts the accepted measurement outcomes to the logical subspace of interest.

\begin{figure}[h]
	\centering
	\includegraphics[width=80mm]{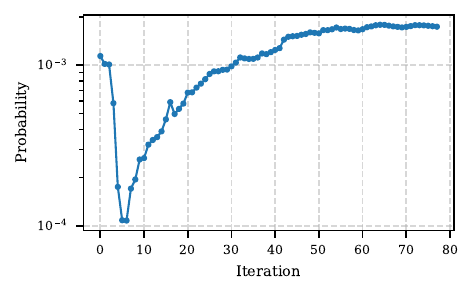}
    \\[4pt]
	\includegraphics[width=80mm]{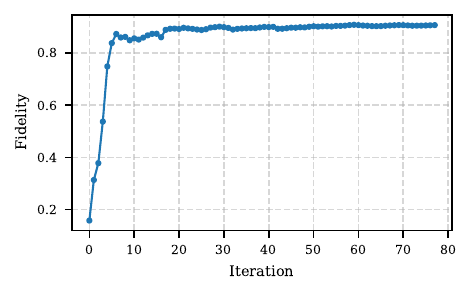}
	\caption{Probability and fidelity of the GHZ state generated with post-selecting non-vacuum states as a function of optimal steps. The weights \(w_1\), \(w_2\) and \(\epsilon\) are set to \(10\), \(1\) and \(10^{-4}\).}
	\label{fig:trace_nv_2}
\end{figure}

It is useful to distinguish between the raw heralding success probability and the effective probability after post-selection. Throughout this subsection, the reported probability refers to the experimentally observable effective success probability, which includes both the heralding probability and the additional non-vacuum post-selection filtering. Although this filtering inevitably reduces the total event rate, it substantially suppresses vacuum-induced errors and thus significantly enhances the fidelity of the generated states.

Fig. ~\ref{fig:trace_nv_1} shows the evolution of the optimized probability and fidelity for a Bell-state target under non-vacuum post-selection. The optimization weights are chosen as \(w_1 = 10\), \(w_2 = 1\), and \(\epsilon = 10^{-4}\), which prioritize the suppression of vacuum contributions during the early stages of the optimization. We have verified that the qualitative improvement is robust against moderate variations of these parameters.

For the Bell-state target, we obtain an optimal effective success probability of \(0.0312\) and a fidelity of \(0.9998\), using only two heralding modes and without photon addition or subtraction. Given the low-gain limitation inherent to probabilistic single-photon sources, this performance is more efficient than that of single-photon-source-based schemes with a comparable number of resources\,\cite{PhysRevResearch.3.043031}. In particular, increasing the number of heralding modes typically leads to an order-of-magnitude reduction in success probability, whereas the present post-selection strategy achieves a high fidelity with only a modest additional filtering cost.

\begin{figure}[h]
	\centering
	\includegraphics[width=80mm]{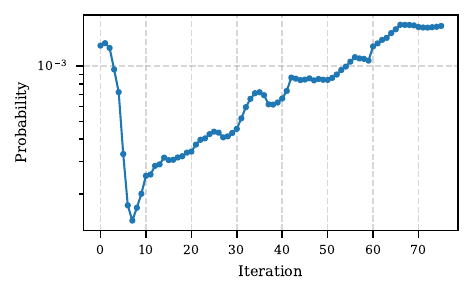}
    \\[4pt]
	\includegraphics[width=80mm]{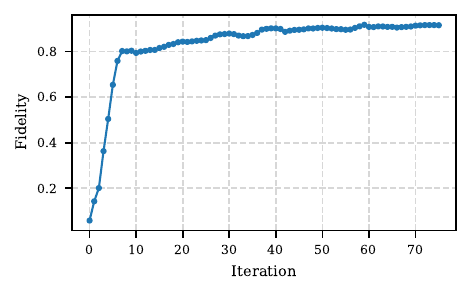}
	\caption{Probability and fidelity of the W state generated with post-selecting non-vacuum states as a function of optimal steps. The weights \(w_1\), \(w_2\) and \(\epsilon\) are set to \(10\), \(1\) and \(10^{-4}\).}
	\label{fig:trace_nv_3}
\end{figure}

We then apply the same post-selection procedure to multipartite entangled states. For the GHZ and W states shown in Fig.~\ref{fig:trace_nv_2} and \ref{fig:trace_nv_3}, we obtain optimal effective success probabilities of \(0.00174\) and \(0.00166\), with corresponding fidelities of \(0.9075\) and \(0.9145\), respectively. The lower fidelities compared to the Bell-state case can be attributed to the increased sensitivity of multipartite entanglement to residual vacuum components and mode correlations. Notably, this approach uses the same number of heralding modes and Gaussian sources as those employed in single-photon-source--based schemes for GHZ-state generation\,\cite{PhysRevA.102.012604}, while achieving competitive performance without relying on deterministic single-photon inputs.

\begin{figure}[h]
	\centering
	\includegraphics[width=80mm]{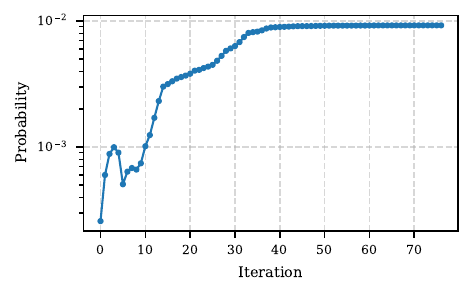}
    \\[4pt]
	\includegraphics[width=80mm]{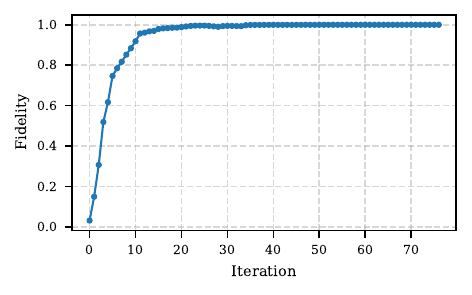}
	\caption{Probability and fidelity of the GHZ state generated with post-selecting non-vacuum states and 2 photons added as a function of optimal steps. The weights \(w_1\), \(w_2\) and \(\epsilon\) are set to \(10\), \(1\) and \(10^{-4}\).}
	\label{fig:GHZ_6_4_2_0_p}
\end{figure}

As shown in the preceding sections, photon addition provides an effective way to improve the fidelity of multipartite entangled states in models without vacuum post-selection. The same strategy can be applied directly in the present post-selected setting. In particular, we find that introducing photon addition during the state-preparation stage remains effective in suppressing residual non-ideal components that are not fully removed by the non-vacuum post-selection alone. 

\begin{figure}[h]
	\centering
	\includegraphics[width=80mm]{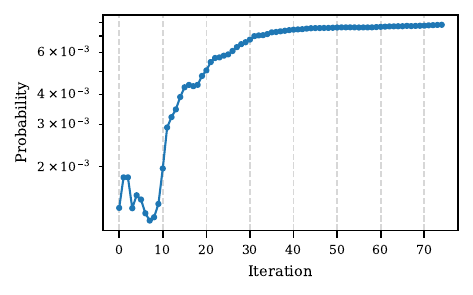}
    \\[4pt]
	\includegraphics[width=80mm]{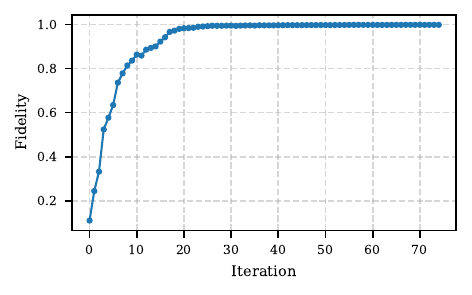}
	\caption{Probability and fidelity of the W state generated with post-selecting non-vacuum states and 2 photons added as a function of optimal steps. The weights \(w_1\), \(w_2\) and \(\epsilon\) are set to \(10\), \(1\) and \(10^{-4}\).}
	\label{fig:W_6_4_2_0_p}
\end{figure}

For both GHZ and W state targets, we find that adding two photons is already sufficient to raise the optimized effective fidelity to values very close to unity. Increasing the number of added photons beyond this point yields only marginal improvement and is therefore not considered. The corresponding optimization trajectories of the effective fidelity and success probability are shown in Fig.~\ref{fig:GHZ_6_4_2_0_p} and Fig.~\ref{fig:W_6_4_2_0_p} under the same non-vacuum post-selection criterion. 

These results demonstrate that the combination of non-vacuum post-selection and a minimal amount of photon addition enables the high-fidelity generation of multipartite entangled states without increasing the number of heralding modes.

\subsection{Robustness}
In realistic experiments, the parameters of Gaussian states cannot usually be set exactly to their optimized values and are subject to fluctuations over time. It is therefore necessary for the model to maintain high generation probability and fidelity under a reasonable level of perturbation. 

\begin{figure}[h]
	\centering
	\includegraphics[width=80mm]{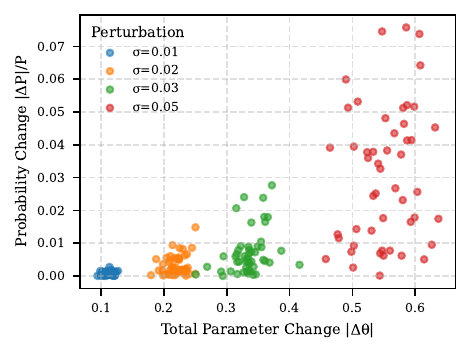}
	\\[4pt]
	\includegraphics[width=80mm]{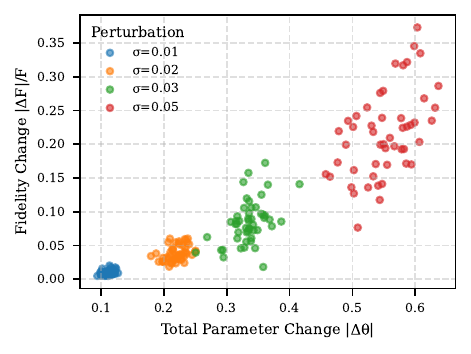}
	\caption{Variations in probability and fidelity as functions of the total parameter change under random perturbations. Different colors indicate the maximum variation applied to each parameter in the random perturbations.}
	\label{fig:robust}
\end{figure}

We assess the robustness of the heralded generation by examining how the probability and fidelity vary with changes in the Gaussian state parameters, which is crucial for its experimental implementation, as it ensures stable performance under parameter fluctuations. 

We introduce random perturbations to each parameter and calculate the differences in the entanglement generation probability and fidelity between the perturbed and optimized results. The results in Fig.~\ref{fig:robust} show that the variations in probability and fidelity follow the same trend as the parameter perturbations. When the fluctuation of each parameter is below \(1\%\), the corresponding changes in probability and fidelity remain within \(0.5\%\) and \(3\%\), respectively. These results indicate that the model exhibits good robustness against small perturbations, with the generation probability being more stable than the fidelity.

\section{Discussion}
In this work, we have optimized a Gaussian-source-based heralded entanglement generation model and presented the optimal generation probabilities and fidelities for dual-rail encoded Bell, GHZ, and W states at the same scale as the single-photon input model. Owing to the inherent vacuum and multiphoton components in Gaussian states, the Gaussian-input scheme exhibits no intrinsic advantage over single-photon inputs, and its performance further degrades in higher-dimensional or multipartite settings. 

By adjusting the heralding modes and incorporating photon addition and subtraction, however, the optimal probabilities and fidelities can be significantly improved. Through optimization under various conditions, we identified Gaussian-state configurations that achieve higher fidelity and generation probability. The introduced non-Gaussian operations further enhance the non-classicality of the output state and improve the efficiency of heralded entanglement generation, consistent with observations reported in other studies of Gaussian-based heralded state generation. 

To further suppress the influence of vacuum and multiphoton components, photon-number-resolving or quantum nondemolition measurements can be applied for post-selection of the desired entangled state, which effectively enhances the output fidelity under low-gain conditions. Post-selecting on the modes associated with each qubit leads to a pronounced improvement in the fidelity. In this regime, the generation probability of the entangled state exhibits a clear advantage over heralded generation schemes based on probabilistic single-photon sources.

To bridge the gap between theoretical models and experimental implementations, we analyzed the robustness of the generation performance under parameter perturbations of the Gaussian sources. The results indicate that when the fluctuations in model parameters are below \(1\%\), the resulting variations in probability and fidelity remain within \(3\%\), demonstrating strong resilience against experimental imperfections.

With the rapid development of Gaussian light sources and integrated photonics, this framework can be readily adapted to realistic experimental platforms. Future work may combine the present optimization approach with specific device parameters to design resource-efficient and experimentally feasible entanglement generation schemes across a variety of photonic architectures.
\begin{acknowledgments}
This work was supported by Quantum Science and Technology-National Science and Technology Major Project (No.\,2021ZD0301200), National Natural Science Foundation of China (Nos.\,12474494, 12204468).
\end{acknowledgments}

\bibliographystyle{apsrev4-2}
\bibliography{refs}

\end{document}